\documentstyle[aps,preprint,epsf]{revtex}
\begin{document}
\draft

\preprint{KIAS-P98025, SNUTP98-108}

\title{A Non-perturbative Evidence toward the Positive Energy
Conjecture for asymptotically locally $AdS_5$ IIB Supergravity
on $S^5$}
\author{Youngjai Kiem$^{(a)}$\footnote{ykiem@kias.re.kr}
           and Dahl Park$^{(b)}$\footnote{dpark@ctp.snu.ac.kr} }
\address{ $^{(a)}$ School of Physics, KIAS, Seoul 130-012, Korea\\
          $^{(b)}$ Center for Theoretical Physics, Seoul National
                   University, Seoul 151-742, Korea}
\maketitle

\begin{abstract}
We consider the classical solution of the type IIB supergravity 
spontaneously compactified on $S^5$, whose metric depends 
only on the radial coordinate and whose asymptotic geometry is 
locally that of $AdS_5$, i.e., $R \times S^1 \times T^2$.   
We solve the equations of motion to obtain the general 
solutions satisfying these conditions, and find that the 
only naked-singularity-free solutions are the $AdS$ black holes 
and $AdS$ solitons.  The other solutions, that smoothly interpolate 
these two solutions, are shown to have naked singularities even 
though their Ricci tensor is proportional to the metric with a 
negative constant. Thus, among the possible solutions of this type, 
the $AdS$ solitons are the unique lowest energy solution; this 
result is consistent with the recently proposed positive energy 
conjecture for the IIB $AdS$ supergravity on $S^5$.  
\end{abstract}

\pacs{04.65.+e, 04.20.Jb, 04.50.+h}

\newpage

\section{Introduction}
The conjectured correspondence between the string theory in
an Anti de Sitter (AdS) space and the conformal field theory 
(CFT) on the boundary opens up a new possibility of studying 
the conformal phase of the Yang-Mills theory via the 
analysis of supergravity, which, in many cases, is a useful 
low energy approximation of the string theory \cite{malda} \cite{gkp} 
\cite{witten}.  From this point of view, one natural case
of interest is the large $N$, near-horizon geometry of
the $D$-threebranes of the type IIB string theory.  In this case,
by the AdS/CFT correspondence, one relates the four-dimensional
supersymmetric Yang-Mills theory on the boundary of the $AdS_5$ to 
the bulk $AdS_5$ IIB supergravity. 

One ultimately hopes to understand the dynamics of the
gauge theory when the supersymmetry of the four-dimensional
Yang-Mills theory is broken.  A practical path toward this goal was 
suggested by Witten \cite{witten2} as follows.
Instead of considering the asymptotically globally $AdS_5$,
one considers the asymptotically locally $AdS_5$ whose asymptotic
geometry allows a non-trivial one-cycle.  On the boundary
Yang-Mills theory side, one assigns the anti-periodic boundary
condition for the fermions around the one-cycle.  This choice
of the boundary condition breaks the supersymmetry by making
relevant scalars and fermions massive, resulting a non-supersymmetric
Yang-Mills theory.  On the gravity side,
in the spirit of the AdS/CFT correspondence, therefore, we are led 
to consider the asymptotically locally $AdS_5$ supergravity
whose asymptotic geometry contains a circle.  In other words,
the asymptotic geometry of the five-dimensional manifolds
that we are interested in should look like $R \times S^1 \times
{\cal M}_2$, where the first $R$ denotes a generic time,
$S^1$ is the circle along which a non-trivial holonomy for 
fermions can be imposed, and ${\cal M}_2$ is a 
two-manifold\footnote{Here,
for simplicity, we assume $S^1$ is 
orthogonal to ${\cal M}_2$.}.  Recalling 
that the original gauge theory contained sixteen supercharges,
we suppose that ${\cal M}_2$ is a Riemann surface.  
We thus consider the 
case when ${\cal M}_2 = M_g$ where $M_g$ is an arbitrary 
Riemann surface with 
genus $g$; on the Yang-Mills theory side, this results the 
(2+1)-dimensional non-supersymmetric Yang-Mills theory in a space-time 
whose space-like hypersurface is a genus $g$ Riemann surface
upon the compactification on the $S^1$.

Horowitz and Myers recently raised the issue of the stability
of the asymptotically locally $AdS_5$ supergravity asking the 
question of validity of the positive energy conjecture \cite{horowitz}.  
When combined with the AdS/CFT correspondence, the instability on the 
supergravity side, if it exists, will either imply the instability of the 
non-supersymmetric Yang-Mills theory or cast some doubts on the exactness 
of the AdS/CFT correspondence.  In fact, the positive energy theorem was 
proved for space-times whose asymptotic geometry is globally $AdS_5$,
consistent with the stability of the supersymmetric Yang-Mills theory
with sixteen supercharges \cite{gibbons}.  However, 
in the context of the asymptotically
locally flat space-time geometries, there are some explicit examples
that show the existence of the regular solutions of the
Einstein equations whose energy is zero \cite{witten3} or not bounded 
from below \cite{brill} \cite{hb}.  In a similar vein, it was 
proved in \cite{lebrun} that the positive energy conjecture fails
for the asymptotically locally Euclidean space-time.  Despite
these aspects of the general relativity with the vanishing cosmological 
constant, the intriguing conjecture of Horowitz and Myers is that the 
negative energy solution of \cite{horowitz}, $AdS_5$ solitons, is the 
lowest energy solution
among the possible asymptotically locally $AdS_5$ solutions
of the $AdS_5$ IIB supergravity, i.e., essentially a positive energy 
conjecture. 
  
In this paper, we find an evidence that supports the Horowitz-Myers 
conjecture; we obtain the general solutions of the IIB supergravity
compactified on a five-sphere whose five-dimensional metric
is of the form $d s^2 = - \alpha (r) dt^2 + \beta (r) ( dx_1^2 + dx_2^2 ) 
 + \gamma(r) d \theta^2 + \delta (r) dr^2 $, which include the  
static genus one sector of the space-times introduced
in the above.  We find that the
AdS solitons (negative energy) of \cite{horowitz} and AdS black holes 
(positive energy) of \cite{witten2} 
are the unique regular solutions whose asymptotic geometry is locally
$AdS_5$, while all other solutions with the same asymptotic 
geometry possess naked singularities that show up in 
the square of the Riemann tensor $R_{\mu \nu \rho \sigma}
R^{\mu \nu \rho \sigma}$. Our results span all the static,
radial-dependent solutions of the genus one sector satisfying
the asymptotic condition; since propagating gravity wave 
degrees of freedom are expected to 
contribute a positive amount to the energy
and since the perturbative analysis of \cite{horowitz}
proves that the $AdS_5$ solitons are local minimum of the
energy, our results suggest that the $AdS_5$ solitons
are the minimum energy solution when we consider the 
space-time geometries whose asymptotic spatial part
of the metric is a product of a circle and a Riemann
surface of genus one, i.e., a two-torus.  In section II,
we start by reviewing the spontaneous compactification
of the IIB supergravity on a five-sphere (see, for example,
\cite{krn} and \cite{duff}), which is the near-horizon
geometry of the $D$-threebrane solutions \cite{malda}.  
In section III, we explicitly solve the equations of motion to get 
the general solutions of the form introduced in the above.  
In section IV, we show that the $AdS_5$ solitons and black holes 
are the unique regular solutions.  In fact, the solution space is
parameterized by points on $R^2$ and, at a fixed radius
of the solution space, the $AdS_5$ solitons and the
$AdS_5$ black holes are two points on a solution space
circle at which naked singularities are absent.  We conclude by 
discussing the cases of an arbitrary genus and,  
especially, the most interesting case of the genus zero
for the test of the Horowitz-Myers conjecture.

\section{Spontaneous compactification of IIB supergravity on $S^5$}

We start from the following action from the IIB supergravity
\begin{equation}
I_{10}= \frac{1}{16 \pi G_{10}}
   \int d^{10}x \sqrt{-g^{(10)}}\left[ e^{-2\phi}
(R^{(10)}+4(D\phi)^2)-\frac{1}{2 \cdot 5!}F^{(10)2} \right] ,
\label{10action}
\end{equation}
where $\phi$ is the dilaton field, $F^{(10)}$ is the RR self-dual 
five-form field strength and $g_{A B}^{(10)}$, the 
ten-dimensional metric.  All other fields are assumed to
be zero.  Here $G_{10}$ is the ten-dimensional Newton's constant.  
We note that when solving the equations of motion,
we have to impose the self-duality condition for the $F^{(10)}$
field by hand \cite{townsend}. In this paper, we consider the 
spontaneous compactification
of the ten-dimensional theory (\ref{10action}) on a $S^5$ \cite{krn}.
For the metric, this consideration amounts to setting it as
\begin{equation}
ds^2=g^{(5)}_{\mu \nu}dx^{\mu}dx^{\nu}+e^{-2\psi}d\Omega^2_5
\label{10metric}
\end{equation}
where the five-dimensional metric $g^{(5)}_{\mu\nu}$ on the $M^5$ and the
radius of the sphere $\exp(-\psi)$ depend only on the coordinates $x^{\mu}$
on the $M^5$. The metric $d\Omega^2_5$ is the standard
metric on the unit $S^5$ with
five angle coordinates $\theta_1$, $\theta_2$, $\theta_3$, $\theta_4$ and
$\theta_5$. Imposing the spherical symmetry on $S^5$ requires that the dilaton
field depends only on the $x^{\mu}$ coordinates. In addition, it is
necessary that the five-form field strength be the scalar on the $S^5$
or be proportional to the pseudoscalar $\epsilon_{\theta_1 \theta_2 \theta_3
\theta_4 \theta_5}$, the volume form on the unit five-sphere. Thus,
the five-form field strength $F^{(10)}$ has nonvanishing components 
$F^{(10)}_{\mu_1 \mu_2 \mu_3 \mu_4 \mu_5}
=F^{(5)}_{\mu_1 \mu_2 \mu_3 \mu_4 \mu_5}$ and 
$F^{(10)}_{\theta_1 \theta_2 \theta_3 \theta_4 \theta_5}=F^{(5)}_1
\epsilon_{\theta_1 \theta_2 \theta_3 \theta_4 \theta_5}$ where $F^{(5)}_1$
is a zero-form field strength on the $M^5$.  
The integration on the $S^5$ after plugging in the fields 
into Eq.~(\ref{10action}) yields
the five-dimensional action 
\begin{equation}
I_5= \frac{1}{16 \pi G_5} \int d^5x \sqrt{-g^{(5)}} 
\left\{ e^{-2\phi_1} \left[ R^{(5)}
+20 e^{2\psi}+4(D\phi_1)^2 -5(D\psi)^2 \right]-\frac{1}{2}e^{5\psi}
(F^{(5)2}_1+F^{(5)2}_2) \right\}
\label{5action}
\end{equation}
where $2\phi_1=5\psi+2\phi$ and $G_5$ is the five-dimensional
Newton's constant.  In deriving Eq.~(\ref{5action}), we 
additionally used an equivalent 
but more convenient description of the five-form field strength $F^{(5)}$ 
by taking the five-dimensional
Hodge dual of the $F^{(5)}$.  Namely, we take 
the Hodge dual transformation via
\begin{equation}
F^{(5)}_2=-\frac{1}{5!}e^{-5\psi} \epsilon^{\mu_1 \mu_2 \mu_3 \mu_4 \mu_5}
F^{(5)}_{\mu_1 \mu_2 \mu_3 \mu_4 \mu_5} \rightarrow
F^{(5)}_{\mu_1 \mu_2 \mu_3 \mu_4 \mu_5}=e^{5\psi} F^{(5)}_2 
\epsilon_{\mu_1 \mu_2 \mu_3 \mu_4 \mu_5}
\label{dualrel}
\end{equation}
where $F^{(5)}_2$ is the zero-form field strength on the $M^5$.
At level of the action, this transformation also dictates that
the term
\[
\frac{1}{16 \pi G_5} \int d^5x \sqrt{-g^{(5)}} 
 \frac{(-1)}{2 \cdot 5!} e^{-5\psi}
F^{(5)2} 
\]
changes to
\[
\frac{1}{16 \pi G_5} \int d^5x \sqrt{-g^{(5)}} 
\frac{(-1)}{2}e^{5\psi}F^{(5)2}_2
\]
to implement Hodge duality, which ensures that the equations of
motion on the $M^5$ are equivalent in both case. We notice that the 
equations of motion for the five-form field strength become the Bianchi
identities in the dual picture. Just as the Bianchi identities for the
five-form field strength in five dimensions are vacuous, the equations
of motion for the zero-form field strength are vacuous.

The Bianchi identities for the zero-form field strengths 
$F^{(5)}_1$ and $F^{(5)}_2$ imply that they are constants. 
The self-duality of the RR five-form field strength furthermore 
gives a constraint $F^{(5) 2}_1 = F^{(5) 2}_2$.
For the spontaneous compactification on the $S^5$ where the
ten-dimensional space-time is the direct product of a five-manifold
and a five-sphere with a constant radius\footnote{It is reasonable to
suppose that this configuration has the lower energy compared to
higher harmonic modes on the $S^5$.}, we require that 
\begin{equation}
D_{\mu} \psi=0.
\label{cond1}
\end{equation}
The equation of motion for $\psi$ from the 
five-dimensional action Eq.~(\ref{5action}) becomes
\begin{equation}
16e^{-2\phi_1}-e^{3\psi}(F^{(5)2}_1+F^{(5)2}_2)=0
\label{eompsi}
\end{equation}
upon using the condition Eq.~(\ref{cond1}).  Eq.~(\ref{eompsi}) implies 
that the field $\phi_1$ should also be a constant, $D_{\mu} \phi_1=0$.
The equations of motion for $\phi_1$ field and the metric 
$g^{(5)}_{\mu\nu}$ then become
\begin{equation}
R^{(5)}+20 e^{2\psi}=0 
\label{eomphi} 
\end{equation}
and 
\begin{equation}
\left( \frac{1}{2}R^{(5)}+6e^{2\psi} \right) g_{\mu \nu} -R_{\mu \nu}=0 ,
\label{eomg}
\end{equation}
respectively under the conditions Eq.~(\ref{cond1}) and 
$D_{\mu} \phi_1 = 0$.  Here $R_{\mu \nu}$ is the Ricci tensor for the metric.
In fact, the equation of motion for the metric, Eq.~(\ref{eomg}),
becomes the equation of motion for $\phi_1$ field, Eq.~(\ref{eomphi}),
if we take the trace of Eq.~(\ref{eomg}); Eq.~(\ref{eomphi})
is a consistent consequence of Eq.~(\ref{eomg}). 
In summary, after the spontaneous compactification on $S^5$,
the equations of motion become Eq. (\ref{cond1}), 
$D_{\mu} \phi_1 = 0$, and Eq.~(\ref{eomg}). 
We are led to the following effective action 
which produces the resulting equations of motion
\begin{equation}
I=\int d^5x \sqrt{-g^{(5)}} \left( R^{(5)}+12 e^{2\psi} \right)
\label{eaction}
\end{equation}
where $\psi$ is constant.  This is the five-dimensional Einstein
equation with a negative cosmological constant.  

\section{Derivation of static solutions} 

The five-dimensional metric that we consider in this
paper, as explained in the section I, is of the form
\begin{equation}
ds^2=g^{(2)}_{\alpha \beta} dx^{\alpha} dx^{\beta} + e^{2\psi_1}
d\theta^2 +e^{\psi_2} (dx_1^2+dx_2^2)
\label{5metric}
\end{equation}
where the two-dimensional metric $g^{(2)}_{\alpha \beta}$ on the 
${\cal M}_2$, 
the radius $\exp (\psi_1)$ of $S^1$ and the size $\exp (\psi_2)$ of 
$T^2$ depend only on the coordinates $x^{\alpha}$ of 
${\cal M}_2$.  
We will eventually consider only static solutions and this means
that the metric fields will depend only on a space-like combination 
of the two coordinates $x^{\alpha}$.  Under these assumptions, similar 
to the $s$-wave reduction of the four-dimensional general relativity 
to a two-dimensional dilaton gravity (see for example \cite{nappi}), 
we can dimensionally reduce the five-dimensional problem to an
equivalent two-dimensional problem.
After rescaling of the metric, $g^{(2)}_{\alpha\beta}=
 e^{-(3\psi_1+\psi_2)/2}g_{\alpha\beta}$, the resulting two-dimensional
action is computed to be 
\begin{equation}
I_2=\int d^2x \sqrt{-g}e^{-2\bar{\psi}}\left[R+
12 e^{2\psi+\bar{\psi}-\psi_1}-\frac{3}{2}(D\psi_1)^2 \right]
\label{2action}
\end{equation}
where $-2\bar{\psi}=\psi_1+\psi_2$.
We choose a conformal gauge for the two-dimensional metric
\begin{equation}
ds^2=g_{\alpha\beta}dx^{\alpha}dx^{\beta}
=-e^{2\rho}dx^+ dx^-.
\label{2metric}
\end{equation}
We get the static equations of motion from the two-dimensional action 
Eq.~(\ref{2action}) under
the choice of the conformal gauge by imposing all fields depend 
only on a spacelike coordinate $x \equiv x^+-x^-$. The static equations 
of motion are then summarized by a one-dimensional action
\begin{equation}
I_1=\int dx \left( \Omega' \rho'+\frac{3}{2}e^{2\psi} e^{-\psi_1}
\Omega^{1/2}e^{2\rho}-\frac{3}{4}\Omega\psi_1^{'2} \right)
\label{1action}
\end{equation}
where $\Omega=e^{-2\bar{\psi}}$ and the prime denotes the 
differentiation with respect to $x$.
The gauge constraint 
\begin{equation}
\Omega''-2\rho'\Omega'+\frac{3}{2}\Omega\psi_1^{'2}=0.
\label{gconst}
\end{equation}
should be supplemented to the equations from Eq.~(\ref{1action}).
From here on, we follow the scheme of \cite{pk} to solve the
equations of motion.  

We find that there are three symmetries of the action (\ref{1action})
\begin{eqnarray*}
&(a)&~ \psi_1 \rightarrow \psi_1 + \alpha,
~\rho \rightarrow \rho +\frac{1}{2}\alpha \\
&(b)&~ x \rightarrow x+\alpha \\
&(c)&~ x \rightarrow e^{\alpha}x,~\Omega \rightarrow e^{\alpha} \Omega,
~\rho \rightarrow \rho -\frac{3}{4}\alpha
\end{eqnarray*}
where $\alpha$ is an arbitrary real parameter of each symmetry 
transformation for
the three fields in our problem. Using these three symmetries we can
integrate the coupled second order differential equations once to
get coupled first order differential equations by constructing
three Noether charges.
They are given by
\begin{eqnarray}
\psi_0&=&3\Omega \psi_1'-\Omega', \label{psi0} \\
c_0&=&\Omega'\rho'-\frac{3}{4}\Omega \psi_1^{'2}-\frac{3}{2}
e^{2\psi}e^{-\psi_1}\Omega^{1/2}e^{2\rho}, \label{c0} \\
s+c_0x&=&\Omega \rho'-\frac{3}{4}\Omega', \label{cs}
\end{eqnarray}
corresponding to each symmetry.
The gauge constraint Eq.~(\ref{gconst}) and the equations of motion
for $\rho$
\begin{equation}
\Omega''=3e^{2\psi}e^{-2\psi_1}\Omega^{1/2}e^{2\rho}
\label{eomrho}
\end{equation}
determine $c_0=0$, and thus the total number of
constants of motion will be reduced from six to five.
To get the general static solutions of Eqs.~(\ref{psi0})-(\ref{cs}), 
we introduce the following field $A$ by
\begin{equation}
QA=\Omega' ,
\label{defA}
\end{equation}
where a positive number $Q$ satisfies $Q^2=3e^{2\psi}$.
Then from Eq.~(\ref{eomrho}) we have
\begin{equation}
Q=e^{\psi_1}\Omega^{-1/2}e^{-2\rho}A' .
\label{chargeQ}
\end{equation}
Using Eqs.~(\ref{psi0}), (\ref{cs}), (\ref{defA}) and 
(\ref{chargeQ}), we get the following equation
\[
\left( \frac{8}{3}QA+2s-\frac{1}{3}\psi_0 \right)A' = Q
\frac{d}{dx} \left( e^{2\rho} e^{-\psi_1} \Omega^{3/2} \right) ,
\]
which can be integrated to yield
\begin{equation}
Q e^{2\rho} e^{-\psi_1} \Omega^{3/2}=\frac{4}{3}QA^2+(2s-\psi_0/3)A+c_1
\equiv P(A) ,
\label{rho}
\end{equation}
where we introduced a function $P(A)$ and $c_1$ is a constant of
integration.
We note that the sign of $P(A)$ should be positive definite.
Combining Eqs.~(\ref{psi0})-(\ref{cs}), (\ref{defA}) 
and (\ref{rho}), we find
\begin{equation}
c_1=-\frac{1}{6}\frac{\psi_0^2}{Q}.
\label{c1}
\end{equation}
Putting Eq.~(\ref{rho}) into Eq.~(\ref{chargeQ}), we get
\begin{equation}
\Omega \frac{dA}{dx}=P(A)~\rightarrow~\Omega\frac{d}{dx}=P(A)
\frac{d}{dA}.
\label{xtoA}
\end{equation}
By changing the differentiation variable from $x$ to $A$ with the help
of Eq.~(\ref{xtoA}) and using Eq.~(\ref{defA}), we 
immediately find that Eq. (\ref{psi0}) can
be integrated to give
\begin{equation}
\psi_1(A)=\frac{1}{8}\ln|P(A)|-\frac{1}{8}(2s-3\psi_0)I(A)+\psi_{10}
\label{psi1}
\end{equation}
where $I(A)=\int P^{-1} dA$ and $\psi_{10}$ is a constant of integration.
In a similar way, we can rewrite Eq.~(\ref{defA}) as
\[
\frac{1}{\Omega} \frac{d\Omega}{dA}=\frac{QA}{P(A)} ,
\]
which, upon integration, becomes
\begin{equation}
\Omega=|P(A)|^{3/8}e^{-(6s-\psi_0)I(A)/8} \Omega_0
\label{omega}
\end{equation}
where $\Omega_0$ is a positive definite constant of integration.
The field $\rho$ can be obtained from Eqs.~(\ref{rho}), (\ref{psi1})
and (\ref{omega})
\begin{equation}
e^{2\rho}=|P(A)|^{9/16}e^{(14s+3\psi_0)I(A)/16}Q^{-1}
e^{\psi_{10}}\Omega_0^{-3/2}.
\label{srho}
\end{equation}
The field $A$ in terms of $x$ can be determined from Eq.~(\ref{xtoA})
\begin{equation}
x-x_0=\int \frac{\Omega(A)}{P(A)} dA ,
\label{coorx}
\end{equation}
where $x_0$ is a constant of integration.
We have solved all the equations of motion; the fields are given by
Eqs.~(\ref{psi1}), (\ref{omega}), (\ref{srho}) and (\ref{coorx})
and there are five constants of motion, $s$, $\psi_0$, $\psi_{10}$, 
$\Omega_0$ and $x_0$.
Since $D \equiv 4s^2-4s\psi_0/3+\psi_0^2$ is positive semi-definite,
which becomes zero only when $s = \psi_0 = 0$, 
the quadratic equation $P(A)=0$ has two 
real roots $A_{\pm}$ where
\[
A_{\pm}=\frac{\psi_0-6s \pm 3\sqrt{D}}{8Q}.
\]
We thus write P(A) as
\[
P(A)=\frac{4}{3}Q(A-A_+)(A-A_-).
\]
Since $Q>0$, we see that $A_+-A_-=3\sqrt{D}/(4Q)>0$.  As can be seen 
from Eq.~(\ref{rho}), we have $P(A)/Q >0$, which restricts
the range of $A$ variable to $A_+<A$ or $A_->A$. 
It turns out that two regions describe the same space-time
if we transform $s \rightarrow -s $ and $\psi_0 \rightarrow
- \psi_0$ (and if we appropriately change the relationship
between $r$ and $A$ below).  Avoiding redundancy, we thus choose 
the region $A_+<A$.
A convenient choice of a radial coordinate $r$ is
\begin{equation}
\frac{r^4}{l^4}=\frac{1}{\sqrt{3}}(A-A_-)
\label{rcoord}
\end{equation}
where $l$ is a constant satisfying $l^2=e^{ -2 \psi}$ and thus
denotes the radius of the $S^5$.
Using this radial coordinate, we can rewrite the ten-dimensional
metric as
\begin{eqnarray}
&&ds^2=\frac{r^2}{l^2}\left(1-\frac{r_0^4}{r^4}
\right)^{(\sqrt{D}-\psi_0-2s)/(4\sqrt{D})}\left[
\left(1-\frac{r_0^4}{r^4}\right)^{\psi_0/\sqrt{D}}d\theta^2
+dx_1^2+dx_2^2
-\left(1-\frac{r_0^4}{r^4}\right)^{2s/\sqrt{D}}dt^2 \right]
\nonumber \\
&&~~~~~~~~~~~~~~~~~~~~~~~~~~~~~~~~~~~~~~~~~~~~~~~~~~~~~~~~~~~~~~~
+\frac{l^2}{r^2}\left(1-\frac{r_0^4}{r^4}\right)^{-1}dr^2
+e^{-2\psi}d\Omega_5^2
\label{smetric}
\end{eqnarray}
where 
\[r_0^4=\frac{\sqrt{3D}}{4Q} l^4,
\]
and we have fixed two constants of integration as
\begin{equation}
e^{4\psi_{10}}=\frac{1}{2\sqrt{Q}},~~~\Omega_0^4=\frac{1}{8Q^{3/2}} ,
\end{equation}
which corresponds to the scale choice for the circle and the two-torus.
Of the original five constants of integration, these two scale choices 
fix two of them.  Since $x_0$ can be deleted from the metric by an 
appropriate diffeomorphism, we can disregard it as a gauge dependent 
parameter. Thus we end up with two constants of integration that 
parameterize the solution space in a gauge-independent fashion.
Hereafter, we will rewrite $\psi_0 = \alpha_1$ and
$2s  = \alpha_2$.

\section{Discussions}

The five-dimensional metric that we obtain from the analysis in
this paper is as follows
\[
ds^2  =  g^{(5)}_{\mu \nu} dx^{\mu} dx^{\nu} = 
       \frac{r^2}{l^2} \left[ 
- \left(1-\frac{r_0^4}{r^4}\right)^{a_t}
  d t^2 
+ \left(1-\frac{r_0^4}{r^4}\right)^{a_1}
  d x_1^2 
+ \left(1-\frac{r_0^4}{r^4}\right)^{a_2}
  d x_2^2 \right. \]
\begin{equation}
\left. + \left(1-\frac{r_0^4}{r^4}\right)^{a_{\theta}}
  d \theta^2  \right]
+ \frac{l^2}{r^2} \left(1-\frac{r_0^4}{r^4}\right)^{-1} dr^2  
\label{general}
\end{equation}
and the ten-dimensional metric\footnote{After the initial
submission of this paper, R. Myers informed us of 
Ref.~\cite{russo}, where the same metric was independently 
obtained.  The analysis in the previous section 
shows that this metric corresponds to the general static, 
radial-dependent solution consistent with the boundary 
conditions that we imposed.} is the tensor product
of the above metric with a five-sphere of a fixed 
radius $l$.
Here $a_t$, $a_1$, $a_2$ and $a_{\theta}$ are given by
\[ a_t = \frac{- \alpha_1 + 3 \alpha_2 + \sqrt{D}}{4 \sqrt{D}} \ , \
   a_1 =  a_2 = \frac{- \alpha_1 - \alpha_2 + \sqrt{D}}{4 \sqrt{D}} \ , \
   a_{\theta} = \frac{3 \alpha_1 - \alpha_2 + \sqrt{D}}{4 \sqrt{D}} , \]
\[  D =  \alpha_1^2 - \frac{2}{3} \alpha_1 \alpha_2 + \alpha_2^2 \]
in terms of $\alpha_1$ and $\alpha_2$; they
satisfy $a_t + a_1 + a_2 + a_{\theta} = 1$ resulting $\det g = - r^6 / l^6$.
Eq.~(\ref{general}) represent the general solutions (under the conditions 
that we imposed earlier) that are asymptotically locally $AdS_5$.  
By direct computation of the Ricci tensor, we can confirm 
that the metric of Eq.~(\ref{general}) satisfies
\begin{equation}
 R_{\mu \nu } = - \frac{4}{l^2} g_{\mu \nu} \ \ \rightarrow \ \
  R = - \frac{20}{l^2}  \ \ , \ \ R_{\mu \nu } R^{\mu \nu } = 
   \frac{80}{l^4}
\end{equation}
for all values of $r_0$, $\alpha_1$ and $\alpha_2$.

For a given value of $~l$, the physically distinct solutions given in 
Eq.~(\ref{general}) are parameterized by points on $R^2$.  To see this,
we first note that the parameter 
$r_0$ plays the role of a radial coordinate\footnote{We note that the
solution does not change as we take $r_0 \rightarrow - r_0$.  Thus,
we concentrate on the non-negative values of $r_0$.}.  When $r_0 = 0$, 
Eq.~(\ref{general}) reduces to the $AdS_5$ space-time regardless of
the values of $\alpha_1$ and $\alpha_2$.  Furthermore, the seemingly
two parameters $\alpha_1$ and $\alpha_2$ combine to give a single angular
coordinate; we observe that the solutions remain invariant under the 
transformation of the constants of motion $\alpha_1$ and $\alpha_2$ via 
$( \alpha_1 , \alpha_2) \rightarrow ( c \alpha_1 , c \alpha_2)$ 
for an arbitrary positive number $c$.  This turns $(\alpha_1 , \alpha_2)$
into a one-dimensional real projective space, i.e., a circle ${\cal S}^1$.

\begin{figure}
\epsfxsize=16cm
\epsfysize=5cm
\epsfbox{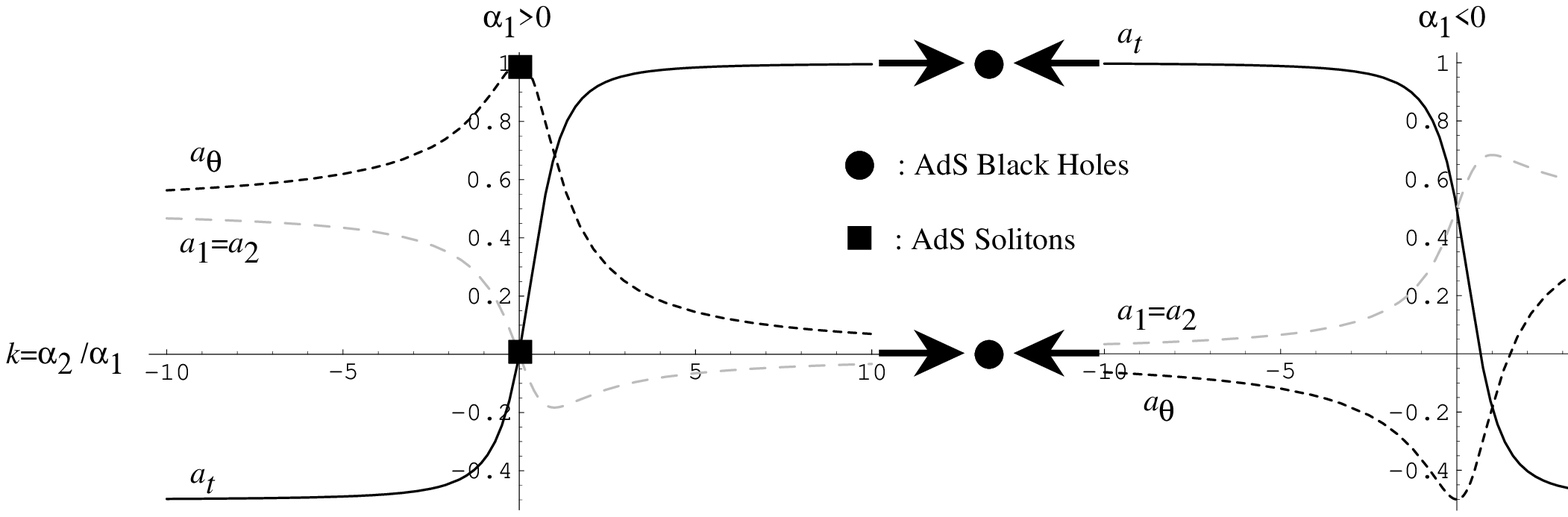}
\caption{The powers $a_t$, $a_1$, $a_2$, $a_{\theta}$ in the metric 
(\ref{general}). The point $k=\pm \infty$ in the left figure is 
idenified with the point $k=\mp \infty$ in the right figure. }
\label{fig12}
\end{figure}

\begin{figure}
\epsfxsize=16cm
\epsfysize=5cm
\epsfbox{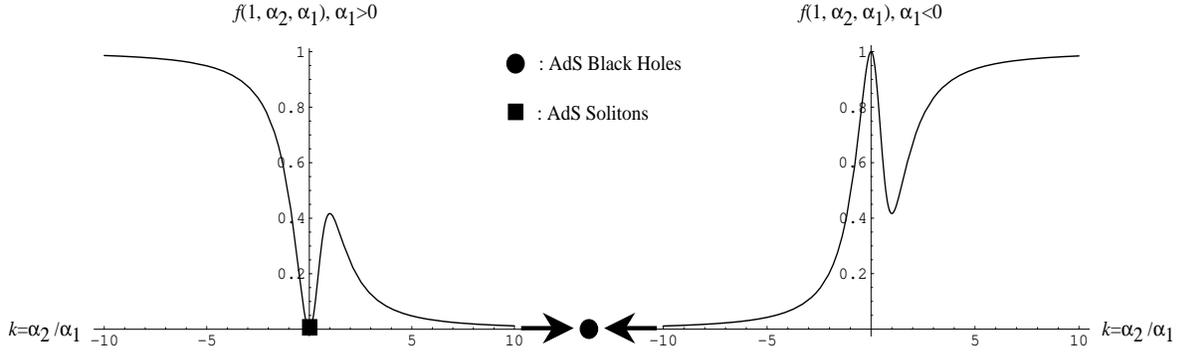}
\caption{The function $f(1,\alpha_1,\alpha_2)$ in (\ref{RR4}).
The point $k=\pm \infty$ in the left figure is idenified with the point 
$k=\mp \infty$ in the right figure.}
\label{fig34}
\end{figure}

We observe the behavior of $a_t$, $a_1$, $a_2$ and $a_{\theta}$
as we traverse along ${\cal S}^1$ for a fixed non-zero value 
of $r_0$ in Fig.~\ref{fig12}.
%
%FIGURE!!!!
%
We recognize two interesting points on ${\cal S}^1$.  The first one
is the $\alpha_1 = 0$ and $\alpha_2 = 1$ case, which corresponds to
the $AdS_5$ black holes.  
\begin{equation}
ds^2  = \frac{r^2}{l^2} \left[
- \left(1-\frac{r_0^4}{r^4}\right) \ d t^2 + d x_1^2 + d x_2^2  
+ d \theta^2  \right]
+ \frac{l^2}{r^2} \left(1-\frac{r_0^4}{r^4}\right)^{-1} dr^2  
\label{adsbh}
\end{equation}
The second is the $\alpha_1  = 1$ and $\alpha_2 =0$ case,
which corresponds to the $AdS_5$ solitons that have negative
energy relative to the $AdS_5$ vacuum \cite{horowitz}.
\begin{equation}
ds^2  =  \frac{r^2}{l^2} \left[ 
- d t^2 + d x_1^2  + d x_2^2  
+ \left(1-\frac{r_0^4}{r^4}\right) d \theta^2  \right]
+ \frac{l^2}{r^2} \left(1-\frac{r_0^4}{r^4}\right)^{-1} dr^2  
\label{adssoliton}
\end{equation}
The metric (\ref{adssoliton}) of the $AdS_5$ solitons
can be obtained from the metric (\ref{adsbh}) of the $AdS_5$ black 
holes by the double Wick rotation \cite{horowitz}.  
In our language, rather similar to the $U(1)$ electric-magnetic
duality, the $\pi /2$ rotation along the ${\cal S}^1$ of the 
black hole metric (\ref{adsbh}) yields the soliton metric 
(\ref{adssoliton}).  In addition, the general solutions 
Eq.~(\ref{general}), that correspond to the rotation along the 
${\cal S}^1$ by an arbitrary angle, reveal the existence of 
``dyonic" solutions, some of which might have, in principle, a lower 
energy than that of the $AdS_5$ soliton (\ref{adssoliton}).  
However, we can show that all of these ``dyonic" solutions have
naked singularities.  For this purpose, we directly compute the
square of the Riemann tensor from the metric (\ref{general})
to find
\begin{equation}
l^4 R_{\mu \nu \rho \sigma} R^{\mu \nu \rho \sigma} 
= 40  + 72 \frac{r_0^8}{r^8}
 + 72  f( r/r_0 , \alpha_1 , \alpha_2 )
 \frac{r_0^{16}}{ r^8 ( r^4 - r_0^4)^2 } ,  
\label{RR4}
\end{equation}
where the function $f(x, \alpha_1 , \alpha_2 )$ is given by
\[ f(x, \alpha_1 , \alpha_2 ) =
\frac{ [ (\alpha_1 - \alpha_2 )^2 ( \alpha_1 + \alpha_2 ) \sqrt{D}
 - D^2  ]  ( 1- 2 x^4 ) - 8 \alpha_1^2 \alpha_2^2 / 27}{2 D^2} . \]
The first term in Eq.~(\ref{RR4}) is the contribution from the
$AdS_5$ vacuum, and the second term is the contribution from the $AdS_5$ 
black hole (or the same contribution from the $AdS_5$ soliton).  We note 
that the third term has the double pole singularity in $r^4$ variable 
at $r_0^4$.  From the spatial infinity ($AdS$ boundary) to the point 
$r=r_0$, the position of the horizon in the case of the $AdS_5$
black holes, the radial space-time evolution is smooth.  Due to the
singularity at $r =  r_0$, however, the space-time geometry of
Eq.~(\ref{general}) has a naked singularity unless, at least, 
$f(1, \alpha_1 , \alpha_2 ) = 0$.  For the lack of singularity
at $r = r_0$, in fact, both coefficients of the $x^4$ term and the 
constant term of $f(x, \alpha_1 , \alpha_2 )$ should vanish since the 
third term of Eq.~(\ref{RR4}) has a double pole singularity.   
The value of $ f(1, \alpha_1 , \alpha_2 ) $, which ranges from
zero to one, as we traverse along the circle ${\cal S}^1$ is
shown in Fig.~\ref{fig34}.
%
%FIGURE!!!!!
%
We immediately find from the figures that there are only two points
on ${\cal S}^1$ at which the function $ f(1, \alpha_1 , \alpha_2 ) $
vanishes.  They precisely correspond to the $AdS_5$ solitons and
the $AdS_5$ black holes.  For these two cases, the whole function
$  f(x, \alpha_1 , \alpha_2 ) $ is identically zero and, as a result,
the third term in Eq.~(\ref{RR4}) itself disappears.  This consideration
shows that all of the general solutions Eq.~(\ref{general}) 
other than the $AdS_5$ black holes and $AdS_5$ solitons have
naked singularity at $r = r_0$.  

To summarize, the general solutions of the five-dimensional IIB 
supergravity spontaneously compactified on a five-sphere 
satisfying the asymptotic condition (asymptotically 
locally $AdS_5$) and of the form 
$ds^2 = - \alpha (r) dt^2 + \beta (r) ( dx_1^2 + dx_2^2 ) 
 + \gamma(r) d \theta^2 + \delta (r) dr^2 $ are parameterized
by all points on $R^2$.  Among these points, naked-singularity-free
solutions are $AdS_5$ black holes (the ray from the origin toward
the positive $x$-axis of $R^2$), $AdS_5$ solitons (the ray from the origin
toward the positive $y$-axis of $R^2$) and the vacuum $AdS_5$ 
space-time (the origin of $R^2$), and all other solutions possess naked 
singularities.  This result points to the feasibility of the part 
of the Horowitz-Myers conjecture, which, in turn, via
the AdS/CFT correspondence, will be related to the 
stability of the (2+1)-dimensional non-supersymmetric Yang-Mills
theory on a space-time whose spatial section is a two-torus.  

For the hermiticity of the gauge theory
Hamiltonian, the fermions on the $S^1$ should either be periodic 
(R sector) or anti-periodic (NS sector).  It was pointed out
in \cite{horowitz} that the negative energy of the $AdS_5$ solitons
is the same as the Casimir energy of the NS sector 
fermions on a circle up to a factor 3/4 (also see \cite{klebanov}). 
In the case of the R sector fermions, the Casimir energy vanishes
and, for the $AdS_5$ black holes, the energy comes
solely from its mass.  The gravity side analysis is consistent
with this picture in a sense that for a given value of $r_0$ there are 
only two discrete points on a circle ${\cal S}^1$ that 
correspond to the naked-singularity-free 
solutions.  Since the invariant size of the Ricci tensor is
bounded from the above (including the point $r =r_0$) for the
interconnecting solutions, one can get the
formal expression for the energy by evaluating the surface
integral Eq.~(2.1) of \cite{horowitz} as follows
\begin{equation}
\frac{E}{{\rm Volume}_{ S^1 \times T^2} } =
- \frac{r_0^4}{16 \pi G_5 l^5} \frac{ \alpha_1 - 3 \alpha_2 }
  { \sqrt{D} } ,
\label{energy?}
\end{equation}
although its interpretation as the true energy is not clear due to the 
presence of the naked singularities of the Riemann tensor at $r = r_0$.  
The quantity Eq.~(\ref{energy?}) 
calculated in this fashion smoothly increases monotonically, 
as we move along the circle ${\cal S}^1$ from the $AdS_5$ solitons
(negative energy $- r_0^4/ (16 \pi G_5 l^5 )$ for $\alpha_1 =1 $ and  
$ \alpha_2 = 0$) to the $AdS_5$ black 
holes (positive energy $ 3r_0^4/(16 \pi G_5 l^5)$ due to the black hole 
mass for $\alpha_1 = 0$ and $\alpha_2 = 1$).    
Pushing the AdS/CFT correspondence further, one
might expect that the solutions that interconnect these
two points and possess naked singularities can be related
to non-unitary gauge theories with
the fermions whose holonomy along the circle $S^1$ multiplies
the fermion field by an arbitrary $U(1)$ phase factor.  It remains
to be seen if this expectation is true.

An interesting generalization of the analysis presented in 
this paper
is to consider the space-time whose asymptotic geometry is of
the form $R \times S^1 \times M_g$ for $g \ne 1$.  In the case
of the higher genus, there will be extra positive contribution 
to the effective potential that results from the negative
curvature of the Riemann surface.  Therefore, we may expect
that the regular solutions in this case will have higher energy
than the energy of the $AdS_5$ solitons.  The situation is more
dangerous for the genus zero case.  As an effective $s$-wave
sector, we may expect that the solutions in this sector, 
{\em if exist}, would
have lower energy than the energy of the $AdS_5$ solitons; in fact,
the explicit examples of \cite{brill} and \cite{hb}, the bubble
type solutions, that 
have an energy which is not bounded from below belong to
the $g = 0$ case with the vanishing cosmological constant.  
More careful analysis of this case will be necessary to 
prove the Horowitz-Myers conjecture and therefore to 
establish the stability of the (2+1)-dimensional non-supersymmetric
Yang-Mills theory on a space-time whose spatial section is 
a sphere.

\acknowledgements{Y.K. would like to thank Seungjoon Hyun and
Hyeonjoon Shin for useful
discussions.  D.P. was supported by the Korea
Science and Engineering Foundation (KOSEF) through the Center for
Theoretical Physics (CTP) at Seoul National University (SNU).}


\end{document}